\documentclass[onecolumn,journal]{IEEEtran}
\usepackage{mathrsfs}
\usepackage{txfonts}
\usepackage [dvips] {graphicx}

\ifCLASSINFOpdf
\else
\fi
\begin{document}
%
\title{A Robust Beamformer Based on Weighted Sparse Constraint}
%
%
%

\author{Yipeng~Liu,
        Qun~Wan,
        and~Xiaoli~Chu
\thanks{Yipeng Liu and Qun Wan are with the Electronic Engineering Department,
University of Electronic Science and Technology of China, Chengdu, 611731 China
e-mail: ({liuyipeng,~wanqun}@uestc.edu.cn).}
\thanks{Xiaoli Chu is with the Division of Engineering, King's College London,
London WC2R 2LS, UK e-mail: (xiaoli.chu@kcl.ac.uk).}
\thanks{Manuscript received Month Day, 2010; revised Month Day, Year.}}

%
%

\markboth{Journal Title,~Vol.~X, No.~X, Month~Year}%
{Shell \MakeLowercase{\textit{et al.}}: Bare Demo of IEEEtran.cls for Journals}
%



\maketitle

\begin{abstract}
Applying a sparse constraint on the beam pattern has been suggested to
suppress the sidelobe level of a minimum variance distortionless response
 (MVDR) beamformer.
In this letter, we introduce a weighted sparse constraint in the beamformer
design to provide a lower sidelobe level and deeper nulls for interference
avoidance, as compared with a conventional MVDR beamformer. The proposed
beamformer also shows improved robustness against the mismatch between
the steering angle and the direction of arrival (DOA) of the desired signal,
caused by imperfect estimation of DOA.
\end{abstract}


%
\IEEEpeerreviewmaketitle

\section{Introduction}
%
%
%
%
\IEEEPARstart{M}{ultiple-antenna} systems have received considerable
 attention from both the wireless industry and academia, because of
 their strong potential in realizing high data rate wireless communications
 in next generation wireless networks. A beamformer is a versatile form of
 spatial filtering, using multiple antenna systems to separate signals that
 have overlapping frequency spectra but originate from different spatial
 locations. Beamforming has become a key technique in current and future
 wireless communications, radar, sonar, etc [1].

 The minimum variance distortionless response (MVDR) beamformer has been
 considered as a popular method for enhancing the signal from the desired
 direction while suppressing all signals arriving from other directions as
 well as the background noise [1], but its relatively high sidelobe level
 would lead to significant performance degradation, especially with unexpected
 increase in interference or background noise [2]. In order to provide sidelobe
 suppression for an MVDR beamformer, a sparse constraint on the beam pattern was
 recently proposed in [3], but this sparse constraint was simply weighted
 equally in every direction, resulting in limited improvement in sidelobe
 suppression. At the same time, possible mismatch between the steering angle
 of the beamformer and the direction of arrival (DOA) of the signal of
 interest (SOI) may further degrade the beamforming performance [4].

 In this letter, we take into account the DOAs of interfering signals in the design
 of a robust beamformer. More specifically, we incorporate the a posteriori DOA
 distribution of interfering signals, which can be coarsely estimated through a
 correlation operation at the beamformer, into the sparse constraint, so that the
 beam pattern is weighted in different directions according to the spatial
 distribution of interference. Numerical evaluations will show that the proposed beamformer
 achieves a much lower sidelobe level, deeper nulls for interference avoidance,
 and stronger robustness against DOA estimation errors, as compared with existing
 designs of beamformers.

\section{MVDR BEAMFORMER}

The signal received by a uniform linear array (ULA) with M antennas can be represented
 by an M-by-1 vector, x(k), the expression of which is given by

\begin{equation}
{\bf{x}}(k) = s(k){\bf{a}}(\theta _0 ) + \sum\limits_{j = 1}^J
{\beta _j (k){\bf{a}}(\theta _j )}  + {\bf{n}}(k)
\end{equation}
where \emph{k} is the index of time, \emph{J} is the number of
interference sources, s(\emph{k}) and $\beta _i (k)$ (for \emph{j} =
1,~...~,~ \emph{J}) are the amplitudes of the SOI and interfering
signals at time instant k, respectively, $\theta _l$ (for \emph{l} =
0, 1,~...~, \emph{J}) are the DOAs of the SOI and interfering
signals,
 $ \varphi _l  = \left( {{{2\pi d} \mathord{\left/
 {\vphantom {{2\pi d} \lambda }} \right.
 \kern-\nulldelimiterspace} \lambda }} \right)\sin \theta _l
$, with \emph{d} being the distance between two adjacent antennas
and $\lambda$ being the operating wavelength [8-10], i.e., the
wavelength of the SOI, and \textbf{n}(\emph{k}) is the additive
white Gaussian noise (AWGN) vector at time instant \emph{k}.

The output of a beamformer for the time instant k is then given by

\begin{equation}
y(k) = {\bf{w}}^H {\bf{x}}(k) = s(k){\bf{w}}^H {\bf{a}}(\theta _0 )
+ \sum\limits_{j = 1}^J {\beta _j (k){\bf{w}}^H {\bf{a}}(\theta _j
)}  + {\bf{w}}^H {\bf{n}}(k)
\end{equation}
where \textbf{w} is the \emph{M}-by-1 complex-valued weighting
vector of the beamformer.

The MVDR beamformer is designed to minimize the total array output
energy, subject to a linear distortionless constraint on the SOI.
The weighting vector of the MVDR beamformer [1] is given by

\begin{equation}
{\bf{w}}_{MVDR}  = \mathop {\arg \min }\limits_{\bf{w}} \left(
{{\bf{w}}^H {\bf{R}}_x {\bf{w}}} \right),{\rm{  s}}{\rm{.t}}{\rm{.
}}{\bf{w}}^H {\bf{a}}(\theta _0 ) = 1
\end{equation}
where ${{\bf{R}}_x }$  is the \emph{M}-by-\emph{M} covariance matrix
of the received signal vector \textbf{x}(\emph{k}), and ${\bf{w}}^H
{\bf{a}}(\theta _0 ) = 1 $ is the distortionless constraint applied
on the SOI.

\section{Weighted Sparse Constraint Beamformer}

In order to suppress the sidelobe level of the conventional MVDR
beamformer, a sparse constraint on the beam pattern was suggested in
[3]. Accordingly, the weighting vector of the improved MVDR
beamformer based on a sparse constraint is given by

\begin{equation}
{\bf{w}}_{SC}  = \mathop {\arg \min }\limits_{\bf{w}} \left(
{{\bf{w}}^H {\bf{R}}_x {\bf{w}} + \gamma \left\| {{\bf{w}}^H
{\bf{A}}} \right\|_p^p } \right),{\rm{  s}}{\rm{.t}}{\rm{.
}}{\bf{w}}^H {\bf{a}}(\theta _0 ) = 1
\end{equation}
where $ \gamma $ is the factor that controls the tradeoff between
the minimum variance constraint on the total array output energy and
the sparse constraint on the beam pattern, the \emph{M}-by-\emph{N}
matrix \textbf{A} consists of \emph{N} steering vectors for all
possible interference with DOA in the range of $ {\rm{[ 90^\circ ,
}}\theta _0 {\rm{)}} \cup {\rm{(}}\theta _0 {\rm{, 90^\circ ]}} $,
with $ \theta _0 $ being the DOA of the SOI as defined in (1), i.e.,
\begin{equation}
{\bf{A}} = \left[ {\begin{array}{*{20}c}
   1 & 1 &  \cdots  & 1  \\
   {\exp \left( {j\varphi _1 } \right)} & {\exp \left( {j\varphi _2 } \right)} &  \cdots  & {\exp \left( {j\varphi _N } \right)}  \\
    \vdots  &  \vdots  &  \ddots  &  \vdots   \\
   {\exp \left( {j\left( {M - 1} \right)\varphi _1 } \right)} & {\exp \left( {j\left( {M - 1} \right)\varphi _2 } \right)} &  \cdots  & {\exp \left( {j\left( {M - 1} \right)\varphi _N } \right)}  \\
\end{array}} \right]
\end{equation}

\begin{equation}
\varphi _l  = \frac{{2\pi d}}{\lambda }\sin \theta _l ,{\rm{ for }}l
= 1, \cdots ,N
\end{equation}
and $\left\| {\bf{x}} \right\|_p  = \left( {\sum\nolimits_i {\left|
{x_i } \right|^p } } \right)^{1/p}$ is is the $\mathscr{C}_p$\ norm
of a vector \textbf{x}.When $0 \le p \le 1 $, the $\mathscr{C}_p$\
norm provides a measurement of sparsity for \textbf{x}. The smaller
the value of $\left\| {\bf{x}} \right\|_p^p $ is, the sparser the
vector \textbf{x} is, meaning that the number of trivial entries in
\textbf{x} is larger [3]. As the product $ {\bf{w}}^H {\bf{A}} $
indicates sidelobe levels of the beam pattern [3], a smaller value
of $ \left\| {{\bf{w}}^H {\bf{A}}} \right\|_p^p $ would imply lower
sidelobe levels. The $\mathscr{C}_p$\ norm can also be explained as
a diversity measure [5].

We consider the \emph{M}-by-1 vector, \textbf{x}(\emph{k}), as a
snapshot of the received signal at time instant \emph{k}. If we
collect the snapshots of \emph{K} ($ K \ge 1 $) different time
instants in a matrix, then we can have an \emph{M}-by-\emph{K} data
matrix as

\begin{equation}
X = \left[ {\begin{array}{*{20}c}
   {{\bf{x}}(1)} & {{\bf{x}}(2)} &  \cdots  & {{\bf{x}}(K)}  \\
\end{array}} \right]
\end{equation}
It has been shown that the cross-correlation of the steering matrix
\textbf{A} with the received data matrix \textbf{X} coarsely
represents the a posteriori spatial distribution of interfering
signals [7]. We use this property to define a weighted sparse
constraint for further suppressing sidelobe levels of the beam
pattern. As a result, the weighting vector of the beamformer with a
weighted sparse constraint is given by
\begin{equation}
{\bf{w}}_{WSC}  = \mathop {\arg \min }\limits_{\bf{w}} \left(
{{\bf{w}}^H {\bf{R}}_x {\bf{w}} + \gamma \left\| {{\bf{w}}^H
{\bf{AQ}}} \right\|_p^p } \right){\rm{, s}}{\rm{.t}}~{\rm{.
}}{\bf{w}}^H {\bf{a}}(\theta _0 ) = 1
\end{equation}
where the \emph{N}-by-\emph{N} matrix \textbf{Q} = diag[SNM($
{\bf{A}}^H {\bf{X}} $)] serves as a weighting matrix, and SNM($
{\bf{A}}^H {\bf{X}} $) is an \emph{N}-by-1 vector containing as
elements the squared normalized mean value of each row of the
\emph{N}-by-\emph{K} matrix $ {\bf{A}}^H {\bf{X}} $ [7]. Comparing
(8) with (4), we can see that the matrix \textbf{Q} in (8) provides
additional weighting on the sparse constraint, in accordance with
the DOA distribution of interfering signals. More specifically, the
larger the probability of interference arriving in a certain
direction, the larger the weight applied on the sparse constraint in
the corresponding direction.

The optimal weighting vector indicated by (8) can be found by using
an adaptive iteration algorithm [3], [5]. When \emph{p} = 1, a
simpler method, called basis pursuit [6], can be used to solve (8)
efficiently. We also observe that (4) can be considered as a special
case of (8), in terms that (8) reduces to (4) when \textbf{Q} =
\textbf{I}, corresponding to the case of equal weighting in every
direction.

In both (4) and (8), the beamformer weighting vector is subject to a
distortionless constraint applied on the SOI. The robust minimum
variance beamforming (RMVB) [10] introduced a modified constraint on
the SOI, for enhancing the beamformer's robustness against mismatch
between the steering angle and the DOA of the SOI. Specifically, the
RMVB weighting vector minimizes the total weighted power output of
the array, subject to the constraint that the array gain exceeds
unity for all array responses within a pre-specified ellipsoid [10],
i.e.,
\begin{equation}
{\bf{w}}_{RMVB}  = \arg \mathop {\min }\limits_{\bf{w}} \left(
{{\bf{w}}^H {\bf{R}}_x {\bf{w}}} \right),{\rm{  s}}{\rm{.t}}{\rm{.
}}~{\mathop{\rm real}\nolimits} \left( {{\bf{w}}^H {\bf{a}}(\theta
)} \right) \ge 1,{\rm{  }}\forall {\bf{a}}(\theta ) \in {\bf{\xi }}
\end{equation}
where real(.) denotes taking the real part, $ {\bf{\xi }} $ is the
ellipsoid that covers all possible values of the estimated
${\bf{a}}\left( {\theta _0 } \right)$, including variations caused
by uncertainty in the array manifold and imprecise knowledge of the
DOA of the SOI [10], and $ {\bf{a}}\left( \theta  \right) $
represents any steering vector that falls within the ellipsoid  . By
comparing (9) with (3), we can see that the RMVB scheme provides an
improvement over the MVDR beamformer, because it explicitly takes
into account variation or uncertainty in the array response, so as
to improve the robustness of the beamformer against mismatch in
steering angle [1], [10].

To improve the robustness against imperfectly estimated steering
angle for our proposed beamformer in (8), we integrate the RMVB as
shown in (9) with the weighted sparse constraint used in (8), and
obtain a robust beamformer as follows

\begin{equation}
{\bf{w}}_{RWSC}  = \mathop {\arg \min }\limits_{\bf{w}} \left(
{{\bf{w}}^H {\bf{R}}_x {\bf{w}} + \gamma \left\| {{\bf{w}}^H
{\bf{AQ}}} \right\|_p^p } \right),{\rm{  s}}{\rm{.t}}{\rm{.
}}~{\mathop{\rm real}\nolimits} \left( {{\bf{w}}^H {\bf{a}}(\theta
)} \right) \ge 1,{\rm{  }}\forall {\bf{a}}(\theta ) \in {\bf{\xi }}
\end{equation}
for which the technique of Lagrange multipliers [1] can be used to
obtain the updating formula of $ {\bf{w}}_{RWSC} $ [3]. The proposed
beamformer in (10) thus includes both a weighted sparse constraint
based on a coarse estimation of interference DOA distribution for
sidelobe suppression and an improved constraint on the SOI that
enhances the beamformer's robustness against DOA estimation errors.

\section{Simulation Results}

In the simulations, an ULA with 8 half-wavelength spaced antennas is
considered. The AWGN noise at each sensor is assumed spatially
uncorrelated. The DOA of the SOI is set to be 0$ ^ \circ $, and DOAs
of three interfering signals are set to be  -30$ ^ \circ $, 30$ ^
\circ $, and 70$ ^ \circ $, respectively. The signal to noise ratio
(SNR) is set at 10 dB, and the interference to noise ratio (INR) is
assumed to be 20 dB, 20 dB, and 40 dB in  -30$ ^ \circ $, 30$ ^
\circ $, and 70$ ^ \circ $, respectively. 100 snapshots were used
for each simulation. Without loss of generality, \emph{p} is set to
be 1 and $ \gamma $ is set to be 2. The matrix \textbf{A} consists
of all steering vectors in the DOA range of $ [ - 90^ \circ  ,0^
\circ  ) \cup (0^ \circ  ,90^ \circ  ] $ with a sampling interval of
1$ ^ \circ $.

Fig. 1 shows beam patterns of an MVDR beamformer, the
sparse-constraint beamformer in (4), and the
weighted-sparse-constraint beamformer in (8). It is obvious that the
best sidelobe suppression performance is achieved by the beamformer
in (8), because the weighted sparse constraint on the beam pattern
explores information about the DOA distribution of interfering
signals. Among the three beamformers, the beamformer in (8) also
provides the deepest nulls in the directions of interference, i.e.,
-30$ ^ \circ $, 30$ ^ \circ $ and 70$ ^ \circ $, especially in 70$ ^
\circ $ where the strongest interference lies.

Fig. 2 shows beam patterns of all beamformers that we have
discussed, with each beamformer having a 3$ ^ \circ $ mismatch
between the steering angle and the DOA of the SOI. We can see that
the MVDR beamformer has a deep notch in 0$ ^ \circ $, which however
is the DOA of the SOI. It can be explained by using the fact that
the MVDR beamformer is designed to minimize the total array output
energy subject to a distortionless constraint on the SOI, so when
the steering angle is in 3$ ^ \circ $, instead of 0$ ^ \circ $, the
MVDR beam pattern maintains distortionless in 3$ ^ \circ $ while
resulting in a deep null in 0$ ^ \circ $. This observation shows the
high sensitivity of the MVDR beamformer to steering angle mismatch.
Comparing beam patterns of beamformers defined in (4), (8), (9) and
(10), we can see that the weighted sparse constraints used in (8)
and (10) are able to further suppress sidelobe levels and deepen the
nulls for interference avoidance, especially in 70$ ^ \circ $, where
the strongest interference locates, as compared with the other two
beamformers. A comparison between the two beamformers in (8) and
(10) shows that the 3$ ^ \circ $ mismatch in the steering angle
leads to an approximate 3$ ^ \circ $ shift of the beam pattern for
the beamformer in (8), while the beamformer in (10) can still
accurately steer the main lobe of the beam patter in 0$ ^ \circ $,
which is the DOA of the SOI. Therefore, our proposed beamformer in
(10) provides significant improvements in terms of sidelobe
suppression, nulling for interference avoidance, and robustness
against DOA estimation errors, with respect to existing beamformers.

\section{Conclusion}
In this letter, we have proposed a robust beamformer based on a weighted
sparse constraint, which leads to a desirable beam pattern with low sidelobe
levels, deep nulls in the directions of strong interference, and an accurate
steering direction in spite of possible estimation error in the DOA of the
desired signal. Numerical experiments have demonstrated that the proposed
beamformer significantly outperforms existing beamformers in terms of sidelobe
suppression, interference avoidance, and robustness against steering angle mismatch.


%

\section*{Acknowledgment}

The authors would like to thank the anonymous reviewers.

This work was supported in part by the National Natural Science Foundation of China under grant 60772146,
the National High Technology Research and Development Program of China (863 Program) under grant 2008AA12Z306
and in part by Science Foundation of Ministry of Education of China under grant 109139.

\ifCLASSOPTIONcaptionsoff
  \newpage
\fi

\begin{figure}[!h]
 \centering
 \includegraphics[scale = 0.47]{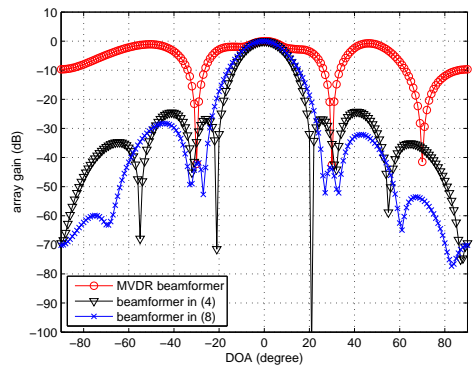}
 \caption{Normalized beam patterns of the MVDR beamformer and beamformers in (4) and (8),
  without mismatch between the steering angle and the DOA of the SOI.}
 \label{figure1}
\end{figure}
\begin{figure}[!h]
 \centering
 \includegraphics[scale = 0.47]{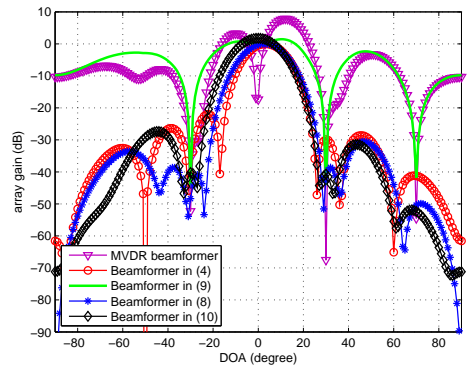}
 \caption{Normalized beam patterns of the MVDR beamformer and beamformers in (4), (8), (9) and (10),
 each having a 3бу mismatch between the steering angle and the DOA of the SOI.}
 \label{figure2}
\end{figure}

\end{document}